\documentclass[preprintnumbers, floatfix,letterpaper, twocolumn,aps,prd,nofootinbib]{revtex4-1}

\usepackage{graphicx}
\usepackage{mathtools, amssymb, amsmath, array}
\usepackage[colorlinks=true, linkcolor=black, urlcolor=blue, citecolor=blue]{hyperref}

\newcommand{\eref}[1]{Equation~(\ref{#1})} 
\newcommand{\Eref}[1]{Equation~(\ref{#1})} 
\newcommand{\sref}[1]{Section~\ref{#1}} 
\newcommand{\fref}[1]{Fig.~\ref{#1}} 
\newcommand{\Fref}[1]{Fig.~\ref{#1}} 

\newcommand{\Cref}[1]{Ref.~\refcite{#1}} 

\newcommand{\aether}{\ae{}ther}
\newcommand{\Aether}{\AE{}ther}

\begin{document}
	
\title{Recovering the Principle of Relativity from the Cosmic Fabric Model of Space}
\author{T. G. Tenev} \email{ticho@tenev.com}
\author{M. F. Horstemeyer} \email{mfhorst@me.msstate.edu}
\date{\today, v1.0}

\begin{abstract}
We extend the descriptive power of the Cosmic Fabric model of space developed by \textcite{Tenev2018} to include moving observers by demonstrating that all reference frames are phenomenologically equivalent with one another and transform between each other via the Lorentz transformations. Our approach is similar to that of~\textcite{Lorentz1892}, which was used to explain the negative outcome of the Michelson-Morley \aether{} detection experiment~\cite{Michelson1887}, except that we deduce the notions of length contraction and time dilation from the postulates of the Cosmic Fabric model. Our result is valid for the continuum length scale at which, by definition, the cosmic fabric can be described mathematically as a continuum. Herein, we also discuss the length-scale dependent nature of the Cosmic Fabric model as a possible way to relate gravitational and quantum theories.

\end{abstract}

\maketitle

\section{Introduction}\label{sec:introduction}

The Cosmic Fabric model proposed by \textcite{Tenev2018} is a formal analogy interpreting space as an elastic solid body (a cosmic fabric) with a Poisson Ratio of unity so only signals in the form of shear waves, such as light and gravitational waves, are admitted. Matter-energy densities behave as inclusions, which are free to move unimpeded within the fabric. These inclusions induce mechanical strain causing the bending of the fabric and the slowing down of clock rates within it, from which effects one can derive the field equations of General Relativity. 

Because of the underlying material analogy, the Cosmic Fabric model is in a sense a descendent from  earlier material models of space, commonly known as \aether{} theories, which have played an indispensable role in our understanding the nature of space. The notion of an \aether{} has been thought about from different perspectives over time probably initially by Aristotle~\cite{RobertCampbell2016} who considered it to be the fifth element comprising the heavenly spheres and bodies. For some time after Aristotle, the \aether{} was viewed as a fluid. For example, although \textcite{Newton1678} described \aether{} as ``capable of contraction and dilatation, strongly elastic,'' which makes one think that he was alluding to a solid \aether{}, he also discussed~\cite{Newton1687} a universal fluid \aether{} filling the cosmos at the largest length scale but a dynamic short range interaction of \aether{} and matter at the smallest length scales. It was \aether{} whose multiple functions admitted transmitted forces to produce the phenomena in the universe that we see, including gravitation.  Later, \textcite{Laplace1796, Laplace1798} furthered the thought that gravity propagated through a liquid \aether{} that had a lower length scale basis from molecules. \textcite{Fresnel1818} proposed that \aether{} was partially entrained by matter, but \textcite{Stokes1845} argued that such entrainment was complete. \textcite{Kelvin1867} described the \aether{} at the smallest length scales as vortex atoms comprising a frictionless, elastic material with ``the hypothesis that space is continuously occupied by an incompressible frictionless liquid acted on by no force, and that material phenomena of every kind depend solely on motions created in this liquid.''  They argued that electromagnetism had to be based on a mechanical notion like that of \aether{}. \textcite{Maxwell1873, Maxwell1878} used the idea of an \aether{} to build the theory of electromagnetic phenomena. \textcite{Lodge1893, Lodge1897} presented mechanical Lagrangian models to illustrate the \aether{}'s phenomenological effects.  

The Lorentz \Aether{} Theory (LET)~\cite{Lorentz1892, Lorentz1895, Lorentz1898} was the culmination of earlier material models of space. Based on Lodge's work~\cite{Lodge1893, Lodge1897}, \textcite{Lorentz1892, Lorentz1895, Lorentz1898, Lorentz1904, Lorentz1921, Michelson1928} developed an electron-\aether{} theory where matter (electrons) and \aether{} were different entities in which the \aether{} was completely motionless. This stationary configuration would not be in motion close to matter. By contrast to earlier electron models, the electromagnetic field of the \aether{} appears as a mediator between the electrons, so a signal cannot propagate faster than the speed of light.  The basic concept of Lorentz's theory~\cite{Lorentz1895} was the ``theorem of corresponding states'' in which an observer moving relative to the \aether{} makes equivalent observations as a stationary observer. \textcite{Lorentz1892} changed the space-time variables between one reference frame and another, and introduced concepts like a physical length contraction and a local time to explain the Michelson and Morley work~\cite{Michelson1881, Michelson1886, Michelson1887}, which had shown that the stationary reference frame of the \aether{} was undetectable. \textcite{Lorentz1898, Lorentz1904} and \textcite{Larmor1897, Larmor1900} discussed that the notion of a local time is accompanied by a time dilation of matter moving in the \aether{}. In other words, there is an elastic \aether{} strain~\cite{Larmor1900} that arises when electrons (matter) are present. \textcite{Larmor1900} tried to view \aether{} in the context of different length scales: electrons, atoms, molecules, and cosmos.  Lorentz later noted~\cite{Lorentz1921, Michelson1928} that he considered a clock stationary in the \aether{} gave the ``true'' time, while local time was thought of as a working hypothesis with a dynamical mathematical backing.  Therefore, Lorentz's theorem is viewed by modern authors as being a mathematical transformation from a ``real'' stationary reference frame in the \aether{} into a dynamic ``fictitious'' configuration. 

Initially, the Theory Special Relativity~\cite{Einstein1905} appeared to have supplanted the material view of space, not because the latter was flawed, but because it seemed no longer needed. However, the more complete understanding that came later when General Relativity~\cite{Einstein1916} reinstated the relevance of attributing material-like nature to space. \textcite{Poincare1900, Poincare1905, Poincare1906} had already introduced the Principle of Relativity in the context of Lorentz's work and declared it as a general law of nature, including gravitation, as he corrected some of Lorentz's mistakes and proved the Lorentz covariance of the electromagnetic equations. He included \aether{} as an undetectable medium that distinguished between apparent and real time. Shortly thereafter, \textcite{Einstein1905} proposed the Special Theory of Relativity in which \aether{} was not necessary, and \textcite{Dirac1951} claimed that \aether{} was ``abandoned'' because of \textcite{Einstein1905}. However, \textcite{Einstein1922a} later came back and discussed the necessity of an \aether{} in the context of the General Theory of Relativity~\cite{Einstein1916}. Lorentz wrote a letter to Einstein in which he speculated that within General Relativity the \aether{} was re-introduced. In his response Einstein wrote that one can in fact speak about a new stationary \aether{} but not a dynamic \aether{}~\cite{Einstein1922, Einstein1924, Einstein1930}.  In a lecture, which \textcite{Einstein1922a} was invited to give at Lorentz's university in Leiden, Einstein sought to reconcile the theory of relativity with Lorentzian aether. In this lecture Einstein stressed that special relativity does not rule out the \aether{}:  
\begin{quotation}
	``To deny the \aether{} is ultimately to assume that empty space has no physical qualities whatever. The fundamental facts of mechanics do not harmonize with this view. For the mechanical behaviour of a corporeal system hovering freely in empty space depends not only on relative positions (distances) and relative velocities, but also on its state of rotation, which physically may be taken as a characteristic not appertaining to the system in itself. In order to be able to look upon the rotation of the system, at least formally, as something real, Newton's objective space. Since he classes his absolute space together with real things, for him rotation relative to an absolute space is also something real. Newton might no less well have called his absolute space `\Aether;' what is essential is merely that besides observable objects, another thing, which is not perceptible, must be looked upon as real, to enable acceleration or rotation to be looked upon as something real.''
\end{quotation}

Although Einstein did eventually affirm the validity of the material view of space, Special Relativity (SR) remained dominant by a wide margin in part because its generalization, that is General Relativity (GR), explained gravitational phenomena better than the Lorentz \Aether{} Theory (LET), which was the state-of-the-art \aether{} theory at the time. In addition, the Principle of Relativity, which is the cornerstone of SR and GR, brought about the unification of space and time into a mathematically elegant spacetime continuum. Nevertheless, the need to separate space from time, such as in the Arnowitt-Deser-Misner (ADM)~\cite{Arnowitt1959, Arnowitt2004} formalism of General Relativity, remained indispensable for GR's practical applications. Although the ADM formalism is not a material model per se, the need for such a formulation points to the continual relevance of a material model of gravity. Such a model, as is LET or the recently introduced Cosmic Fabric model~\cite{Tenev2018} necessarily separates space from time, because space is viewed as a material object progressing in time. \textcite{Tenev2018} laid the ground work for a modern material model of space and discussed the benefits of such a view, which seeks to leverage a century of advancement in Solid Mechanics since the time of LET.

Whereas in~\cite{Tenev2018} we showed that the Cosmic Fabric model is a valid analogy to General Relativity for nearly static observers, herein we extend the analogy to include moving observers as well. Although the Cosmic Fabric Model ostensibly defines a hyperplane of absolute simultaneity and therefore a preferred reference frame, we demonstrate that to both moving and stationary observers alike, such a preferred reference frame remains undetectable at continuum length scales. In the discussion section, \sref{sec:discussion}, we mention briefly how the Cosmic Fabric model is conducive to generalizations outside of the continuum length scale. The essential contribution of this paper is to expand the descriptive power of the Cosmic Fabric Model~\cite{Tenev2018} to include moving observers and show that all reference frames are phenomenologically equivalent with one another, which is to say that the Principle of Relativity applies within the Cosmic Fabric model. Consequently, we conclude that the model has at least the descriptive power of SR and LET, and also, based on~\cite{Tenev2018}, the descriptive power of GR for the case of weak gravity. Therefore, the work herein contributes toward a material model of space that generalizes LET similarly to how GR generalized SR.

Our paper is organized as follows: In \sref{sec:speed-of-signals-invariance}, we show that the speed of signal propagation within the fabric, which we had previously identified with the speed of light in free space, is invariant for all observers within all reference frames in the fabric. This result recovers the Second Postulate of Special Relativity, which states that the speed of light is invariant for all inertial reference frames. In \sref{sec:lorentz-transformations}, we derive the transformations between the coordinates of a stationary and a moving observer, which we recognize as the Lorentz transformations~\cite{Lorentz1892, Lorentz1895, Larmor1897, Larmor1900}. Because these apply in like manner between any two inertial reference frames and not only between a stationary and a moving reference frame, it follows that no reference frame can be singled out as special by an observer within the fabric. Consequently, we recover the First Postulate of Special Relativity~\cite{Einstein1905}, which states that all physical laws are the same in all inertial reference frames. In \sref{sec:discussion} we discuss how the Cosmic Fabric model compares with Special Relativity and Lorentz \Aether{} Theory, and we conclude in \sref{sec:conclusion}.

\section{Invariance of the Speed of Signals}\label{sec:speed-of-signals-invariance}

In this section, we show that the speed of signal propagation is invariant with respect to any inertial reference frame for any observer within the fabric. First, we demonstrate the invariance for stationary observers at different locations. Next, we show that the invariance also applies for a moving and a stationary observer at a given location. The combination of these two results leads to the desired conclusion about the speed of a signal's invariance. 

\subsection{Stationary Observers at Different Locations}

\begin{figure}
	\centering
	\includegraphics[width=0.9\linewidth]{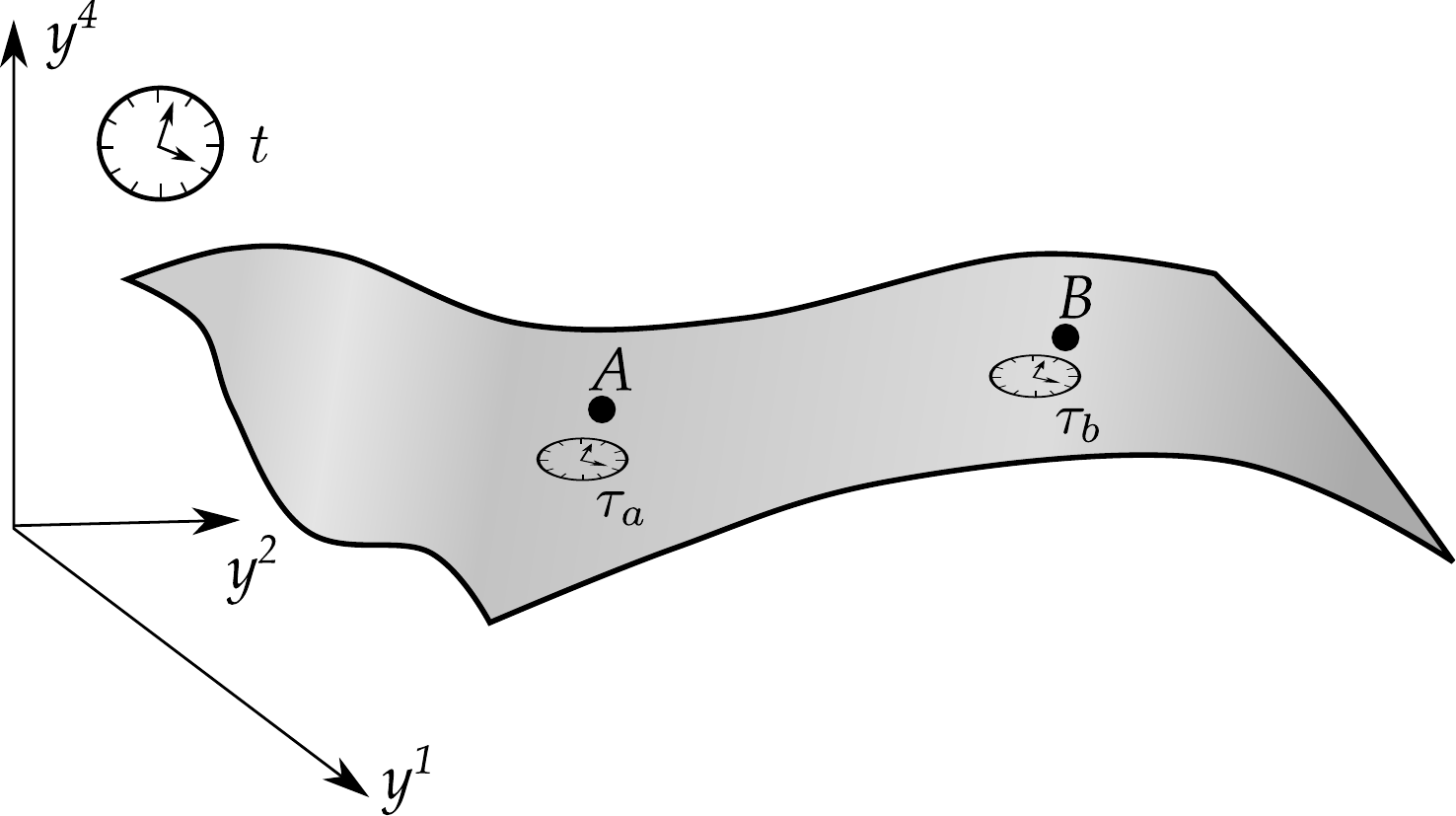}
	\caption{Stationary observers $A$ and $B$ located within the cosmic fabric and measuring proper times $\tau_a$ and $\tau_b$, respectively. The fabric is considered immersed in a four-dimensional hyperspace (not necessarily physical) with coordinates $(t, y^1, y^2, y^3, y^4)$. The third spatial dimension is suppressed for clarity.}
	\label{fig:stationary-observers}
\end{figure}

The formulation of the Cosmic Fabric model~\cite{Tenev2018} postulates that the fabric mediates all matter-matter interactions via signals that travel as mechanical disturbances within it. It is convenient to consider the fabric as immersed within a four-dimensional reference hyperspace (not necessarily physical), which also has its own time coordinate (see \Fref{fig:stationary-observers}). Such a reference space is somewhat similar to Dicke's ``Newtonian coordinate system''~\cite{Dicke1957}, except in the context of four spatial dimensions. From the perspective of this reference space, the rate of matter-matter interactions is proportional to the speed of signal propagation. Consequently, the rate of clock ticks, that is the time lapse rate, is also proportional to the speed of signal propagation.

Although signals need not propagate with uniform speed when measured with respect to the enclosing reference space, nevertheless their speed will appear constant from the perspective of observers within the fabric for the following reason: A clock placed where fabric signals propagate relatively slow in relation to the enclosing reference space will also tick slow. Therefore, to an observer within the fabric, the signal speed would appear to have remained unchanged, because the slowdown of the clock used to measure the signal speed would exactly compensate for the reduction of said speed. Consequently, the signal speed within the fabric appears to be invariant for any stationary observer. In the description of the Cosmic Fabric model~\cite{Tenev2018}, we had identified the speed of signal propagation with the speed of light in free space $c$. Below we provide an algebraic derivation of this result. 

Let $v_a$ and $v_b$ be, respectively, the speeds of signal propagation at locations $A$ and $B$ of the fabric measured in relation to the enclosing reference space. By definition, $v_a = dl_a/dt$, where $dl_a$ is a distance element at location $A$, and $dt$ is the travel time, reckoned with respect to the enclosing reference space.  According to the Time Lapse postulate of the Cosmic Fabric model~\cite{Tenev2018}, $d\tau_a/dt = \left(\frac{1}{v_0}\right) v_a$, where $\left(\frac{1}{v_0}\right)$ is a constant of proportionality. Therefore, the speed of signal propagation $c_a$ measured at location $A$ within the fabric is, 
\begin{equation}\label{eq:c_a}
c_a = \frac{dl_a}{d\tau_a} = \frac{dl_a}{dt}\frac{dt}{d\tau_a} = v_a\frac{v_0}{v_a} = v_0.
\end{equation}

In the same way, we can show that the speed of signal propagation $c_b$ at location $B$ is $c_b = v_0$. Thus, we conclude that the speed of signal propagation at both locations $A$ and $B$ is one and the same with respect to stationary observers in the fabric. The magnitude of this speed can be identified with the speed of light in free space $c$:
\begin{equation}\label{eq:c_a-2}
c_a = c_b = v_0 = c.
\end{equation}

Since locations $A$ and $B$ were arbitrary, it follows that the speed of signal propagation is invariant for all stationary observers within the fabric.

\subsection{Moving Observer}

\begin{figure}
	\centering
	\includegraphics[width=0.9\linewidth]{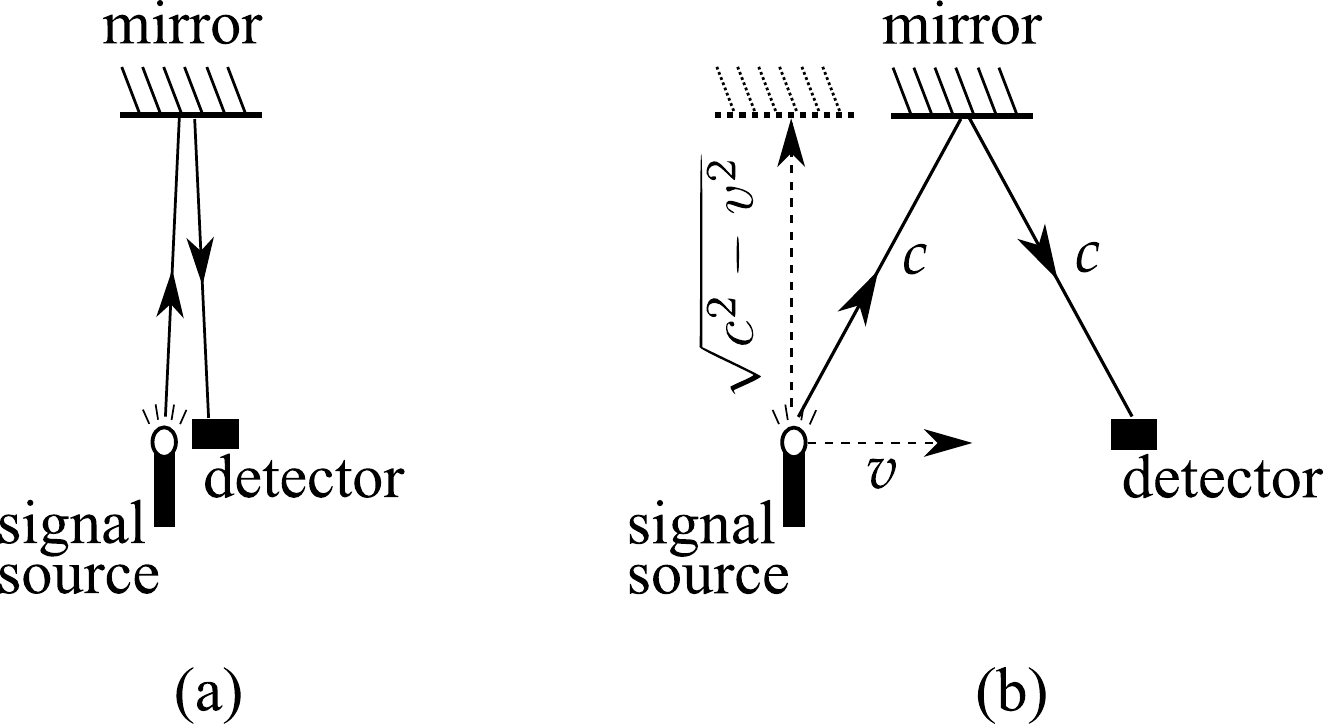}
	\caption{An observer moving with velocity $v$ with respect the fabric. The observer measures the round-trip time of a light signal traveling a fixed distance in a direction transverse to the motion. The situation is represented from the view point of the moving observer (a) and a stationary observer (b), respectively. The effective transverse signal propagation speed is $\sqrt{c^2 - v^2}$. }
	\label{fig:moving-observer}
\end{figure}

We now consider the situation of an observer at a given location moving with velocity $v$ with respect to the fabric (see \Fref{fig:moving-observer}). Let $\tau'$ and $\tau$ represent the time measured by the moving observer and a stationary observer at the same location, respectively. As illustrated on \Fref{fig:moving-observer}, from the Pythagorean Theorem follows that the effective signal speed in any orientation transverse to the motion of the observer is $\sqrt{c^2 - v^2}$. Therefore, by the Time Lapse postulate of the Cosmic Fabric model we conclude that the moving observer's clocks must tick slower by a factor of $(\sqrt{c^2 - v^2})/c = \sqrt{1 - \beta^2}$ where $\beta \equiv v/c$. In other words, 
\begin{equation}\label{eq:time-dilation}
\frac{d\tau'}{d\tau} = \sqrt{1 - \beta^2},\quad \beta \equiv \frac{v}{c}
\end{equation}
We considered the signal speed in a transverse orientation to avoid any direction-specific effects, such as length contraction.

\begin{figure}
	\centering
	\includegraphics[width=0.6\linewidth]{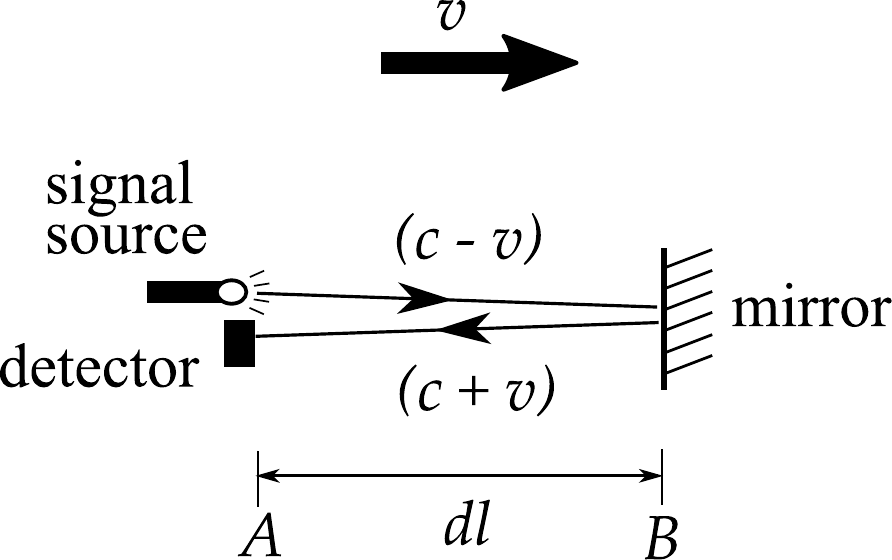}
	\caption{ A rod $AB$ with rest length $dl$, moving with velocity $v$, and aligned along the orientation of motion. A comoving observer is measuring the rod by timing the round-trip signal sent from one end of the rod to the other and back. The situation is represented from the view point of a stationary observer. }
	\label{fig:moving-observer-2}
\end{figure}

Next we consider how lengths are affected along the orientation of motion. We measure lengths by the round-trip time of signals (see \fref{fig:moving-observer-2}). In all of our thought experiments we use round-trip times
to avoid complications due to clock synchronization. Let $dl$ be the rest length of a rod oriented along the direction of motion. From the perspective of the stationary observer, the total time $d\tau$ it takes for the signal to travel from one end of the rod and back is as follows,
\begin{equation}\label{eq:dt}
d\tau = \frac{dl}{c + v} + \frac{dl}{c - v} = 2\frac{dl}{c}\frac{1}{1-\beta^2}.
\end{equation}
From the view point of the moving observer, per \Eref{eq:time-dilation} the round-trip travel time $d\tau'$ is as follows,
\begin{equation}\label{eq:dt-prime}
d\tau' = dt\sqrt{1 - \beta^2} = 2\frac{dl}{c}\frac{1}{\sqrt{1 - \beta^2}}
\end{equation}
The above equation can be interpreted in one of two ways: either the moving observer perceives the round-trip signal speed to differ in the direction of motion compared to any transverse direction, or the rod's length changed such that the new length $dl'$ is $dl' = dl/\sqrt{1-\beta^2}$. The former possibility implies that signal speed, and therefore light speed, would be anisotropic. The Michelson-Morley experiment~\cite{Michelson1887} was designed to measure such anisotropy, and its negative outcome rules out the former possibility. Therefore, one must conclude that measuring rods oriented along the motion of the reference frame experience length change per, 
\begin{equation}\label{eq:dl-prime}
dl' = \frac{dl}{\sqrt{1 - \beta^2}}.
\end{equation}

Due to the combined effects of time dilation and length contraction, the signal speed $c$ is measured to be one and the same by both the moving and stationary observers. Thus, 
\begin{equation}\label{eq:c-invariance}
\frac{dl}{d\tau} = \frac{dl'}{d\tau'} = c
\end{equation}

The above result is the same as what \textcite{Lorentz1892, Lorentz1895, Lorentz1898} concluded in constructing his \aether{} theory, except he considered only the speed of light as opposed to a more general speed of signals as per the Cosmic Fabric model. In the Cosmic Fabric model, the speed of signal propagation is more fundamental than the speed of light, because it controls all matter-matter interactions and not just those pertaining to electromagnetic phenomena. This is why, the variation of speed of signal propagation affects not only the rate of clock ticks but also the length of measuring rods.

\section{Lorentz Transformations}\label{sec:lorentz-transformations}

Next, we recover the Lorentz transformations from the speed of light invariance and from basic considerations of spatial symmetry. The method, is detailed below. 

\begin{figure}
	\centering
	\includegraphics[width=0.9\linewidth]{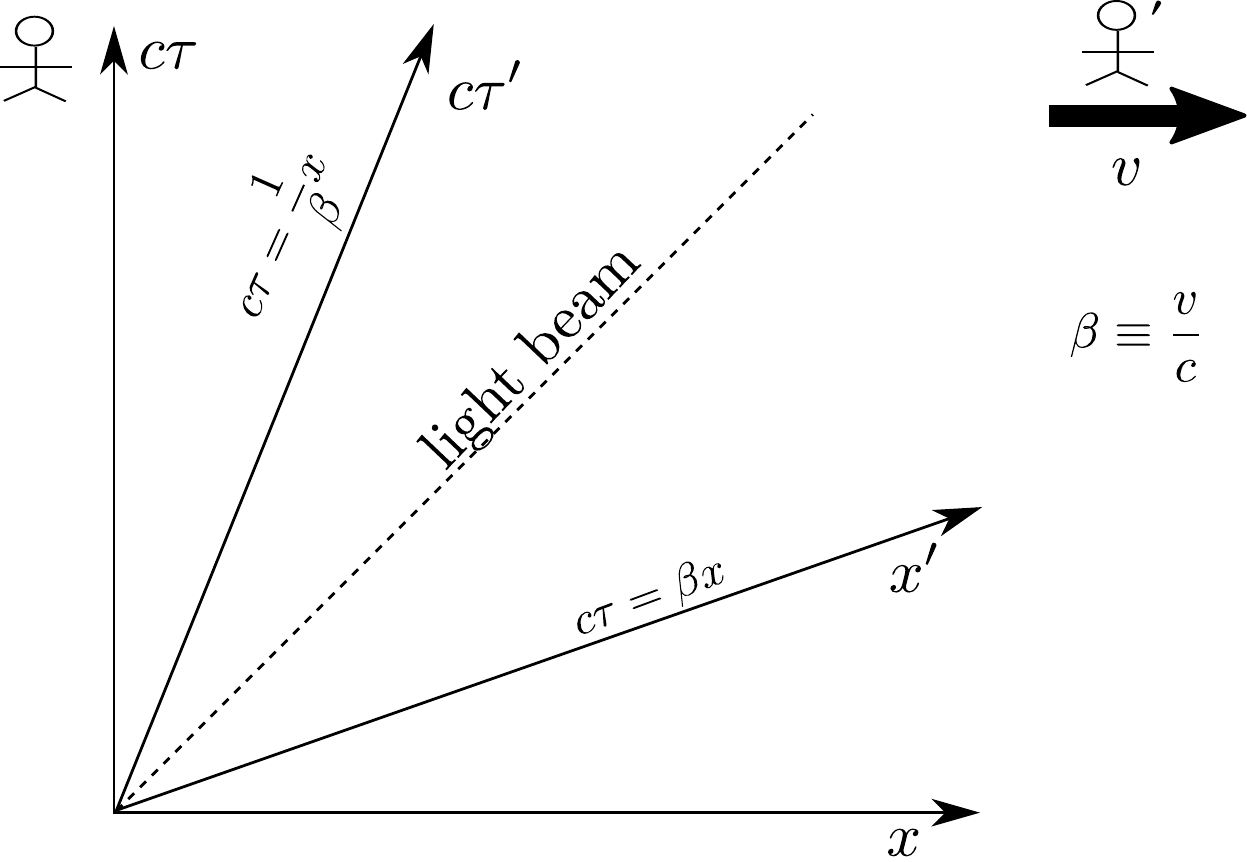}
	\caption{A stationary (unprimed) and a moving observer (primed) represented, respectively, by two set of coordinate axis: $(x, c\tau)$ and $(x', c\tau')$. The primed observer moves with velocity $v$ in the positive $x$ direction. The time dimension is scaled by the speed of light $c$, so that the path traversed by a photon bisects the unprimed axes. Due to the light speed invariance, the same photon path must also bisect the primed axes. For clarity, only one spatial dimension is shown, but the other two are implied. }
	\label{fig:lorentz-transform}
\end{figure}

\Fref{fig:lorentz-transform} shows the spacetime coordinates of an unprimed and primed observer, where the primed observer travels with velocity $v$ with respect to the unprimed one. Let $\beta \equiv v/c$, be the scaled relative speed. The origin represents an event when $x = x' = 0$ and $c\tau = c\tau' = 0$. Because the time coordinates are scaled by $c$, a photon emitted at the origin in the positive $x$ direction will traverse a straight line trajectory that bisects the angle between the unprimed axes. A particle stationary at the primed origin traverses a straight line trajectory of slope $1/\beta$ with respect to the unprimed coordinates. Therefore, the primed time axis must have a slope $1/\beta$ with respect to the unprimed coordinates. Because of the speed of light invariance, the trajectory of the aforementioned photon must also bisect the angle between the primed axes, and consequently, the primed space axis must have a slope of $\beta$ with respect to the unprimed coordinates.

Consider a sufficiently small region around the origin where the cosmic fabric is translationally symmetric in space and time. Such translational symmetry is only applicable at the continuum length scale, at which discrete substructure can be ignored or homogenously included into the fabric's continuous smoothness. The translational symmetry implies that the coordinate transformations we seek must be linear, so let: 
\begin{equation}\label{eq:lorentz1}
\left(\begin{array}{c} x' \\ c\tau' \end{array} \right) = 
\left(\begin{array}{cc} a_{11} &  a_{12} \\ a_{21} & a_{22} \end{array}\right)
\left(\begin{array}{c} x \\ c\tau\end{array} \right) 
\end{equation}
where the coefficients $a_{\mu\nu}$ only depend on the relative velocity $v$. Events along the primed time axis ($x'= 0$) must have unprimed coordinates 
such that $c\tau = x/\beta$. Therefore, $a_{12} = -\beta a_{11}$. Likewise, events along the unprimed time axes ($x = 0$) must have primed coordinates such that $x' = -\beta c\tau'$, and hence $a_{12} = -\beta a_{22}$. 
Finally, events along the primed spatial axis ($c\tau' = 0$) must have unprimed coordinates such that $c\tau = \beta x$, from where we conclude that $a_{21} = - \beta a_{22}$. Consequently, letting $\gamma = a_{11} = a_{22}$, \eref{eq:lorentz1} becomes the following,
\begin{equation}\label{eq:lorentz2}
\left(\begin{array}{c} x' \\ c\tau' \end{array} \right) = 
\gamma \left(\begin{array}{cc} 1 &  -\beta \\ -\beta & 1 \end{array}\right)
\left(\begin{array}{c} x \\ c\tau\end{array} \right) 
\end{equation}
The inverse of the transformation in \Eref{eq:lorentz2} should also have the same form and correspond to equal and opposite velocity, which leads to the following requirement:
\begin{equation}\label{eq:lorentz3}
\begin{split}
\left(\begin{array}{cc} 1 &  0 \\ 0 & 1 \end{array}\right) & = 
\gamma^2 \left(\begin{array}{cc} 1 &  -\beta \\ -\beta & 1 \end{array}\right)
\left(\begin{array}{cc} 1 &  \beta \\ \beta & 1 \end{array}\right) \\
& = \gamma^2 \left(\begin{array}{cc} 1 - \beta^2 & 0 \\ 0 & 1 - \beta^2 \end{array}\right)
\end{split}
\end{equation}
Therefore, we deduce that coefficients $a_{\mu\nu}$ must be as follows,
\begin{equation}\label{eq:lorentz4}
\begin{split}
a_{11} &= a_{22} = \gamma, \quad a_{12} = a_{21} = -\beta = -v/c \\
\gamma & = \frac{1}{\sqrt{1-\beta^2}} = \frac{1}{\sqrt{1-v^2/c^2}}
\end{split}
\end{equation}
which are in fact the coefficients of the well-known Lorentz transformations.

So far we have derived the coordinate transformations between the stationary and a moving reference frames. It is straightforward to verify that the composition of two Lorentz transformations corresponding to scaled velocities $\beta_1$ and $\beta_2$ is also a Lorentz transformation corresponding to scaled velocity $\beta = (\beta_1 + \beta_2)/(1+\beta_1 \beta_2)$. Since a transformation between any two arbitrary reference frames can be treated as a transformation from the one to the special rest frame and from the rest frame to the other, therefore coordinate transformations between any two arbitrary reference frames is also a Lorentz transformation. Because all reference frames within the fabric transform between each other in like manner, it will be impossible to distinguish which one is the special rest reference frame. Thus, we have shown that the Principle of Relativity, which is the first postulate of the Theory of Special Relativity, can be deduced from the postulates of the Cosmic Fabric Model.

\section{Discussion}\label{sec:discussion}

As pointed out in the introduction, the Lorentz \Aether{} Theory (LET)~\cite{Lorentz1892} and Einstein's Special Relativity (SR)~\cite{Einstein1905} are mathematically equivalent. In a sense, LET describes  reality from the perspective of the fabric's enclosing hyperspace, while SR describes the same reality from the perspective of an observer within the fabric. The phenomenological equivalence of LET and SR with the Cosmic Fabric model is limited to the continuum length scale and to the absence of gravity. Indeed, both LET and SR are continuum theories, which means that they treat physical space as smooth and governed by local laws; that is, laws unaffected by any global attributes of space. Furthermore, both LET and SR view physical space as flat (Euclidean), and spacetime also as flat (Minkowskian). The latter limitation is removed by the Theory of General Relativity (GR), which extends SR to curved spacetime. In this sense, one can view the Cosmic Fabric model as extending LET to account for \aether{} curvature and time dilation in a way similar to how GR extends SR. Nevertheless, like SR, General Relativity remains strictly a continuum theory, whereas the Cosmic Fabric model includes parameters, such as its thickness, texture, and an inherent (undeformed) shape, that captures information about the structure of physical space at length scales below and above the continuum scale. 

The derivation of the Lorentz transformations in \sref{sec:lorentz-transformations} depended on the translational symmetry of the fabric, which only holds true at a continuum length scale. At lower length scales, the substructure of the fabric, namely its weave-like composition and its thickness become significant and break the translational symmetry. The weave-like composition of the fabric, which was inferred from its Poisson Ratio being unity, has been discussed in~\cite{Tenev2018}. Examples of materials with a Poisson Ratio of unity include those of \textcite{Rodney2015} and \textcite{Baughman2016}, all of which have complex fibrous substructure.  Symmetry can also be broken by structure at length scales above the continuum, such as at the length scale of the visible universe. Whenever symmetry is broken, the special rest reference frame of the cosmic fabric becomes physically detectable. 

For example, red-shift observations of the Cosmic Microwave Background (CMB)~\cite{Penzias1965} can be used to define a kind of universal rest frame. For any point in space, there is exactly one inertial reference frame within which the CMB redshift appears isotropic. Since the redshift is associated with the expansion of cosmic space, this special reference frame represents an observer that is ``attached'' to space and moves along with it as it expands. Except for this special reference frame, within all other reference fames, the CMB redshift will appear to have a directional dipole indicating the direction and relative velocity of the observer. For example, the dipole we observe from Earth suggests, that the Solar System is moving at about 600 km/s in the direction of the constellation Centaurus~\cite{Kogut1993, Aghanim2014}. The existence of the special CMB rest reference frame means that while local physical observations remain independent of the observer's velocity and position, at the same time cosmic scale observations will differ according to the observer's relative motion in relation to the CMB.

Just like a special rest frame can be identified at cosmological length scales, we speculate that, in a similar way, a special rest frame could be identified at sub-continuum length scales. For example, if space itself has a complex topological structure as \textcite{Misner1973} suggest, then at sufficiently small length scales, such as quantum length scales, it can no longer be modeled as a continuum. That is why a quantum theory of gravity will need to offer a way for bridging between the continuum and sub-continuum length scales. As discussed in \textcite{Horstemeyer2012, Horstemeyer2018}, the field of Solid Mechanics has developed techniques for bridging between length scales within a material. The Cosmic Fabric model, which treats space as a solid material body, provides a way to leverage such existing techniques in the effort for developing a quantum theory of gravity.

\section{Conclusion}\label{sec:conclusion}

Herein we demonstrated that the material analogy of space, as previously introduced by~\textcite{Tenev2018} for the case of nearly static observers, also applies for observers in motion. In the context of continuum length scale, the two postulates of Special Relativity were deduced from the postulates of the Cosmic Fabric model~\cite{Tenev2018}, and all reference frames, whether stationary or moving, were shown to transform between each other using the Lorentz transformations. Therefore, at the continuum length scale, the special rest reference frame remains indistinguishable from any other reference frame and the Principle of Relativity is recovered.
	
At a continuum length scale, the Cosmic Fabric model is to the Lorentz \Aether{} Theory what General Relativity is to Special Relativity. At other length scales, the Cosmic Fabrjic model predicts that there be a physically detectable special rest reference frame. This prediction is consistent with observations of the CMB and likely pertinent in the development of a quantum gravity theory.

\bibliography{recovering-relativity}
	
\end{document}